\documentclass{aa}
\usepackage{txfonts, graphicx, tocbibind}
\usepackage[round]{natbib}
\newcommand{\te}{\ensuremath{T_{\mathrm{eff}}}}
\newcommand{\kms}{km s$^{-1}$}
\newcommand{\vsi}{\ensuremath{v \sin i}}
\newcommand{\bz}{\ensuremath{\langle B_z \rangle}}
\newcommand{\bs}{\ensuremath{\langle B \rangle}}
\begin{document}
\titlerunning{Magnetic Field and Abundance of HD~318107}
\title{Magnetic Field and Atmospheric Chemical Abundances of the Magnetic Ap Star HD~318107\thanks{Based on observations made with the European Southern Observatory telescopes under ESO programmes 065.I-0644(A), 075.C-0234(B), 079.C-0170(A), 081.C-0034(A), and 082.C-0308(A),obtained from the ESO/ST-ECF Science Archive Facility.}}
\author{J. D. Bailey\inst{1}\thanks{Based on observations obtained at the Canada-France-Hawaii Telescope (CFHT) which is operated by the National Research Council of Canada, the Institut National des Science de l'Univers of the Centre National de la Recherche Scientifique of France, and the University of Hawaii.}
\and J. D. Landstreet\inst{1,2}
\and S. Bagnulo\inst{2}
\and L. Fossati\inst{3}
\and O. Kochukhov\inst{4}
\and C. Paladini\inst{5}
\and J. Silvester\inst{6}
\and G. Wade\inst{6}}
\institute{Department of Physics \& Astronomy, The University of Western Ontario, London, Ontario, N6A 3K7, Canada
\and
Armagh Observatory, College Hill, Armagh, Northern Ireland
\and
Department of Physics \& Astronomy, Open University, Milton Keynes, UK
\and
Department of Astronomy \& Space Physics, University of Uppsala, Uppsala, Sweden
\and
Institut f\"ur Astronomie, Universit\"{a}t Wien, T\"{u}rkenschanzstrasse 17, 1180 Wien, Austria
\and
Department of Physics, Royal Military College of Canada, Kingston, Ontario, Canada K7K 7B4}
\date{Received 8 August 2011~/~Accepted 27 September 2011}

\abstract {A new generation of powerful and efficient spectropolarimeters  has recently been used to provide the first sample of magnetic Ap stars of accurately known ages. Modelling of these data offer the possibility of significant new insights into the physics and main sequence evolution of these remarkable stars. }{New spectra have been obtained with the ESPaDOnS spectropolarimeter, and are supplemented with unpolarised spectra from the ESO  UVES, UVES-FLAMES, and HARPS spectrographs, of the very peculiar large-field magnetic Ap star HD~318107, a member of the open cluster NGC~6405 and thus a star with a well-determined age. The available data provide sufficient material with which to re-analyse the first-order model of the magnetic field geometry and to derive chemical abundances of Si, Ti, Fe, Nd, Pr, Mg, Cr, Mn, O, and Ca.} {The models were obtained using ZEEMAN, a program which synthesises spectral line profiles for stars that have magnetic fields.  The magnetic field structure was modelled with a low-order colinear multipole expansion, using coefficients derived from the observed variations of the field strength with rotation phase. The abundances of several elements were determined using spectral synthesis. After experiments with a very simple model of uniform abundance on each of three rings of equal width in co-latitude and symmetric about the assumed magnetic axis, we decided to model the spectra assuming uniform abundances of each element over the stellar surface.}{The new magnetic field measurements allow us to refine the rotation period of HD~318107 to $P = 9.7088 \pm 0.0007$~days. Appropriate magnetic field model parameters were found that very coarsely describe the (apparently rather complex) field moment variations.  Spectrum synthesis leads to the derivation of mean abundances for the elements Mg, Si, Ca, Ti, Cr, Fe, Nd, and Pr. All of these elements except for Mg and Ca are strongly overabundant compared to the solar abundance ratios. There is considerable evidence of non-uniformity, for example in the different values of \bz\ found using lines of different elements.}{The present data set, while limited, is nevertheless sufficient to provide a useful first-order assessment of  both the magnetic and surface abundance properties of HD~318107, making it one of the very few magnetic Ap stars of well-known age for which both of these properties have been studied.}

\keywords{stars:magnetic fields - stars: chemically peculiar}
\maketitle

\section{Introduction}

The existence of main sequence A- and B-type stars with strong magnetic fields has been known for decades.  These peculiar stars (referred to as Ap stars) show anomalous abundances of particular elements, for example of Cr, which may be as much as $10^2$ times (2 dex) overabundant compared to the Sun.  It is also not uncommon to find atmospheric abundances of rare earths well in excess of solar  \citep{Ryabchikova91}.  Almost all Ap stars have angular momenta that are of the order of 10\% or less of typical values for normal stars of similar mass.  The magnetic field strength, spectral line strengths and shapes, and brightness in various photometric bands all vary with the rotation period of the star.  This variability is explained using the oblique rotator model: the magnetic field axis and rotation axis of the Ap star are not aligned with the line of sight, nor with one another, and several elements are distributed non-uniformly over the surface in a pattern that is not axisymmetric about the rotation axis.  This means that observations through the rotation cycle lead to varying field measurements as a result of observing the magnetic field from different aspects. Similarly the spectrum, and (as a result of line blocking and backwarming) photometric magnitudes and colours, vary as different parts of the star are observed \citep{Ryabchikova91}.  

HD~318107 (=NGC 6405 77), a very peculiar magnetic Ap star, has an effective temperature $\te = 11\,800 \pm 500$~K, luminosity $\log L/L_\odot = 1.92 \pm 0.1$, and mass $M/M_\odot = 2.95 \pm 0.15$ \citep{paper2}. It has a large global magnetic field; the mean line-of-sight magnetic field component \bz\ is sometimes as large as 5 or 6~kG, and the typical size of the mean field modulus \bs\ is about 15~kG. The star is a member of the open cluster NGC 6405, so that it is known to have an age of $\log t = 7.80 \pm 0.15$~(yr). This means that the star has completed about $17 \pm 7$\% of its main sequence lifetime \citep{paper2,paper3}. It is still a rather young star.

Initially, a rotation period of 52.4 days was proposed by \citet{North1987}; however, more extensive photometric data obtained by \citet{MM2000} combined with measurements of the mean surface magnetic field \bs\ led to the conclusion that the rotation period is $P = 9.7085 \pm 0.0021$ days.  

The magnetic field of this star was modelled by \citet{LM2000} based on data from \citet{Mathys97}, \citet{MH1997} and further unpublished magnetic field measurements by Mathys. \citet{LM2000} adopted a model of colinear magnetic dipole, quadrupole, and octopole components of strength +23700, -23600, and +8300 G respectively.  Observations of the hemispherically averaged line of sight component of the magnetic field \bz\ indicate that \bz\ is always positive, which leads to the model constraint that the sum of the inclination of the rotation axis to the line of sight, $i$, and the angle between the magnetic field axis and the rotation axis, $\beta$, must be less than about 90 degrees: $i + \beta \lesssim  90^{\circ}$.  This implies that the negative magnetic pole is never directly observed.  For this  model, the values for $i$ and $\beta$ were chosen to be $11^{\circ}$ and $78^{\circ}$, with uncertainties discussed in the text of \citet{LM2000}.  Note that this model does not reproduce the observations very well.  It appears that the field structure of HD~318107 is more complex than the axisymmetric low order multipole field geometry used, and is probably not even axisymmetric.  Nevertheless, the model provides a first order approximation to the magnetic field geometry of HD~318107.

There are rather few hot Ap stars with strong magnetic fields for which detailed chemical abundance analyses have been carried out.  HD~318107, with $\te = 11800$~K, measured mean field modulus of $\bs \sim 15$~kG \citep{Mathys97} and $v \sin i \approx 7$~km~s$^{-1}$, is an excellent object for which to carry out such an analysis.  The low $v \sin i$ of the star together with the large field strength make it possible to observe resolved Zeeman splitting in many spectral lines.  New measurements of \bs\ and \bz\ are reported here, with which it is possible to improve the precision of the rotation period and re-examine the magnetic model. 

HD~318107 is also interesting in that it has an age known with relatively high precision. The large sample of magnetic Ap stars in open clusters for which magnetic fields have been measured and characterised  by \citet{paper1} and \citet{paper2, paper3} have already provided the first clear evidence of evolution of magnetic field strength of the fossil fields of middle main sequence stars with time (or stellar age) during the main sequence phase of evolution. This data set, much of which is based on relatively high-dispersion (polarised) spectra, also offers the possibility of studying empirically the evolution with time and stellar age of the chemical peculiarities of magnetic Ap stars. This is now being undertaken; HD~318107 is the first star from this new sample to be analysed for its chemical abundance and distribution properties. 

This paper will discuss our efforts to model the magnetic field and the chemical abundance distributions of several elements for HD~318107.  The following section discusses the spectroscopic and spectropolarimetric observations used and the field strength values deduced from them;  Sect. 3 discusses improvement of the precision of the period; Sect. 4 describes the modelling technique and the spectral line synthesis program used; Sect. 5 discusses the abundance models obtained; and Sect. 6 summarizes the work presented.

\section{Observations}

\subsection{New data}

\begin{center}
\begin{table*}[ht]
\caption{The new spectroscopic and polarimetric data used in our analysis.  Recorded are the HJD and derived phases for each spectra.  The values of \bz\ for the ESPaDOnS spectra are shown followed by \bs\ values for all spectra measured from three spectral lines: Fe~{\sc ii} $\lambda$6149, Nd~{\sc iii} $\lambda$5050 and $\lambda$6145. }
\centering
\begin{tabular}{lccccccccc}
\hline\hline
Instrument & HJD & Phase & SNR & \bz & \multicolumn{3}{c}{\bs\ (G)} & Window & RV \\
 & & & & (G) & Fe~{\sc ii} $\lambda$6149 & Nd~{\sc iii} $\lambda$5050 & Nd~{\sc iii} $\lambda$6145 & (\AA) & (\kms)\\
\hline
ESPaDOnS & 2454553.119 & 0.991 & 250 & $4900 \pm 45$ & $13900 \pm 380$ & $12600 \pm 340$ & $12400 \pm 280$ & 3690-10481 & $-9$\\
ESPaDOnS & 2454553.150 & 0.995 & 250 & $4856 \pm 42$ & & & & 3690-10481 & $-9$\\
ESPaDOnS & 2454652.793 & 0.258 & 320 & $1985 \pm 31$ & $15400 \pm 380$ & $16300 \pm 340$ & $16800 \pm 280$ & 3690-10481 & $-9$\\
HARPS & 2453463.898 & 0.803 & 80 & & $13600 \pm 420$ & $14900 \pm 410$ & $15100 \pm 420$ & 3783-6914 & $-9$\\
HARPS & 2453582.645 & 0.034 & 90 & & $13600 \pm 500$ & $11900 \pm 680$ & $12000 \pm 670$ & 3783-6914 & $-9$\\
HARPS & 2453583.693 & 0.142 & 40 & & $13900 \pm 420$ & $14100 \pm 550$ & $14200 \pm 570$ & 3783-6914 & $-9$\\
HARPS & 2454223.642 & 0.056 & 70 & & $13200 \pm 310$ & $12300 \pm 510$ & $13000 \pm 410$  & 3783-6914 & $-9$\\
HARPS & 2454633.613 & 0.283 & 55 & & $14700 \pm 250$ & $16400 \pm 410$ & $16400 \pm 340$  & 3783-6914 & $-8$\\
HARPS & 2454869.850 & 0.615 & 70 & & $13800 \pm 420$ & $15700 \pm 680$ & $15300 \pm 570$  & 3783-6914 & $-8$\\
HARPS & 2454870.879 & 0.721 & 60 & & $13600 \pm 420$ & $14700 \pm 680$ & $15900 \pm 570$ & 3783-6914 & $-8$\\
UVES & 2451692.894 & 0.390 & 220 &  &  $15900 \pm 210$ & $17000 \pm 280$  & $16400 \pm 380$ &  4795-6745  \\
FLAMES--UVES & 2454256.625 & 0.453 & 150 &  & $16500 \pm 340$ & $16400 \pm 510$  & $16900 \pm 420$ & 4200-6200 & $-8$\\
\hline\hline
\label{spectra}
\end{tabular}
\end{table*}
\end{center}

Previous observations of HD~318107 include  photoelectric photometry reported by \citet{North1987} and by \citet{MM2000} and used for determination of the rotation period; a series of (unpolarised) spectra obtained by \citet{Mathys97} from which the mean field modulus \bs\ was measured, and used by \citet{MM2000} to redetermine the rotation period; and a small number of spectropolarimetric  observations reported by \citet{MH1997} from which the mean longitudinal field \bz\ was measured.  There is also one new measurement of \bz\ using ESO's FORS1, reported by \citet{paper1}.

The spectra used in this project were obtained more recently than the observations noted above, and come from a variety of sources.  The data available for spectral line synthesis consist of three newly acquired ESPaDOnS spectra and a new FLAMES-UVES spectrum \citep{flames-ref}, together with seven HARPS spectra and one UVES spectrum from the ESO archive.  

ESPaDOnS  is a cross-dispersed echelle spectrolarimeter located at the Canada-France-Hawaii telescope, which can measure all the Stokes parameters ($I, Q, U, V$). We have three flux ($I$) and circular polarisation ($V$) spectra which cover almost the entire window from 3700~\AA\ to 1.04~$\mu$m with a resolving power of $R \simeq 65\,000$.  These spectra typically have signal-to-noise ratios ($SNR$) of the order of 200 over much of this window. Two of the spectra were obtained one after the other, and thus are at essentially the same rotational phase. The $I$ spectra were used for measuring the mean field modulus, and for modelling the chemical element abundance distribution over the stellar surface. \bz\ was measured from the $I$ and $V$ spectra and used to constrain the magnetic model. 

HARPS is located at the European Southern Observatory (ESO) La Silla 3.6 m telescope.  It is a cross-dispersed echelle spectrograph which covers a spectral range of 3780 - 6910~\AA\ with $R = 115\,000$.  The available spectra have $SNR$ between about 40 and 100.  Since the available HARPS spectra do not contain any circular polarisation information, these data were  used for measurement of \bs\  for improvement of the rotation period, and for abundance modelling of HD~318107. 

UVES is a high resolution, cross-dispersed optical spectrograph located at ESO's Paranal Observatory. One spectrum was obtained using the FLAMES instrument, with a resolution of $R = 47\,000$ in a spectral window covering the ranges 4200 to 5160~\AA\ and 5230 to  6200~\AA.  This spectrum was used to measure the mean field modulus \bs. The other spectrum was obtained with the standard UVES red arm setting at 580~nm, using image slicer {\#}3, resulting in a resolving power of 110\,000.

The new data are summarised in Table~\ref{spectra}. The first column reports the instrument used, and the second gives the Julian Date at the middle of the observation. The meaning of the remaining columns will be discussed in more detail below.

The radial velocity of HD~318107 during  each observation was measured either by fitting an appropriate model spectrum to that single observed spectrum during the process of modelling the chemical abundance distributions (see below), or by measuring the wavelengths of a few spectral lines of low Zeeman sensitivity. The measured radial velocity varies somewhat between different methods of measuring it, as the line profiles of individual lines are (often strongly) distorted by Zeeman splitting or by blends. However, the various methods usually differ by no more than about 1~\kms, so we conservatively estimate the uncertainty of radial velocities to be about $\pm 2$~\kms. The measured value for each spectrum, rounded to the nearest 1~\kms,  is reported in the final column, labelled RV, of Table~\ref{spectra}.  The mean radial velocity of the star is $-8.8$~ \kms, and there is no hint in our data, which were taken over a time span of more than eight years, of any variability.  This result is consistent with the mean radial velocity of NGC~6405 (about $-7$~\kms) as reported by \citet{Kharetal05}, and supports the membership of HD~318107 in this cluster. However, our measurements are all quite different from that reported by \citet{MM2000} who measure the radial velocity to be $+9.4 \pm 2.0$~\kms.  This discrepancy may be the result of a sign error by Manfroid \& Mathys, as our new radial velocities show no indication of variability.

\subsection{Longitudinal field strength measurements}

The new ESPaDOnS $V$ and $I$ spectra provide measurements of both \bz\ and \bs. The \bz\ values were obtained by using the first-order moment of the high signal-to-noise $V$ line profile created by the technique of Least Squares Deconvolution (LSD). The method used is discussed in detail in \citet{paper3}. The line list initially used for the LSD processing was a generic list of about 2900 lines suitable for an Ap star of $\te = 12\,000$~K, containing primarily lines of the iron peak elements Ti, Cr and Fe, together with some lines of lighter elements such as Mg and Si. The integration limits were chosen by eye for each spectrum. The resulting measurements are listed in the fifth column of Table~\ref{spectra}. The stated uncertainties were obtained by propagating errors from the underlying $V$ and $I$ LSD spectra. 

In addition to the three measurements reported here, one other new measurement of \bz\ is available from \citet{paper1}. This measurement was obtained using FORS1, and separate values are reported there for the \bz\ value obtained using only the Balmer lines ($+6519 \pm 55$~G), and the value from the metallic spectrum ($+3784 \pm 59$~G). It is obvious that these two measurements agree poorly with one another. This is a phenomenon that is found in several of the measurements reported by \citet{paper1} (for NGC~2169-12, NGC~2244-334, HD~149277, CD~-48~11051, and HD~318107). In each case the star has a \bz\ field of several kG, and all have \te\ values above about 11\,500~K. Because the field values are very large, there is no question about the validity of the detections, but the \bz\ values measured in different ways can differ by as much as a factor of two. Why this kind of discrepancy occurs for some (but not all) large-field stars is not yet understood. It might be connected with the failure of the weak-field approximation which is used to deduce \bz\ values from FORS spectra. However, it may simply be the consequence of magnetic field vectors which vary greatly over the visible hemisphere, possibly even on a fairly small scale, and which are sampled differently when averaged using different chemical elements, which in general are not distributed uniformly over the stellar surface. 

To explore the possibility that inconsistent field measurements are due mainly to the inhomogeneous structure of the magnetic field and the surface distribution of various elements, we have remeasured the longitudinal field \bz\ in our three ESPaDOnS spectra using spectral lines of only a single element at a time. There are enough lines of Si, Ti,  Cr and Fe to provide LSD field measurements of high enough accuracy to distinguish differences at the few hundred G level. The results of this experiment are listed in Table~\ref{Bz-per-el}. 

\begin{table}[ht]
\caption{Measurements of \bz\ using LSD masks composed of lines of single elements (Si, Ti, Cr, and Fe) from the three ESPaDOnS spectra (cf Table~\ref{spectra}).}
\begin{tabular}{cccccc}
\hline\hline
JD               &   \bz(Si)   &   \bz(Ti)   &   \bz(Cr)   &   \bz(Fe)  \\
$- 2450000$  &     (G)     &     (G)      &      (G)      &   (G)        \\
\hline
4553.119  & $7031 \pm 379$ & $4193 \pm 231$ & $4215 \pm 139$ & $5430 \pm 50$ \\
4553.150  & $7107 \pm 370$ & $4222 \pm 221$ & $4124 \pm 132$ & $5237 \pm 47$ \\
4652.793  & $2007 \pm 269$ & $708 \pm 148$ & $878 \pm 92$ & $2417 \pm 32$ \\
\hline\hline
\label{Bz-per-el}
\end{tabular}
\end{table}

From Table~\ref{Bz-per-el} a number of conclusions may be drawn. First, the very satisfactory agreement of the individual entries in the top two lines, derived from two independent spectra obtained within an hour of one another indicates that the claimed precision (measured by repeatability) of the field measurements is believable. On the other hand, in each of the ESPaDOnS spectra, the values of \bz\ measured using different elements reveal remarkable differences. Cr and Ti seem to sample the field in similar ways; both near phase 0 and phase 0.25 the measured \bz\ values are not significantly different. Fe, however, seems to sample more strongly regions of field lines parallel to the line of sight than the other two iron peak elements, and this tendency is still stronger in the sampling of the field associated with Si, which, however,  does not sample the field in the same way as Fe. 

The measurements of the field using different elements thus reveal exactly the same general phenomenon already found in the comparison of the \bz\ value obtained using lines of H to the values obtained using metal (primarily Fe) lines in the single field measurement with FORS1. These results strongly suggest inhomogeneous distributions of various elements, perhaps combined with significant field structure variation on a small or medium scale, have large enough amplitude to greatly affect field measurements. 

\subsection{Field modulus measurements}

The ESPaDOnS, HARPS and FLAMES-UVES spectra all show spectral lines with resolved Zeeman splitting.  We have made measurements of the mean field modulus \bs, which we have compared to those of \citet{MM2000} in order to improve the precision of the rotation period of HD~318107.  The measurements were made using the splitting observed in the Fe~{\sc ii} $\lambda 6149$ line, and in Nd~{\sc iii} $\lambda 5050$ and $\lambda 6145$.  The Fe~{\sc ii} $\lambda 6149$ and Nd~{\sc iii} $\lambda 6145$ lines were used by \citet{MM2000} to measure \bs\ and so to permit direct comparison are used in this study as well.  In addition to these two lines, we also measured the Nd~{\sc iii} $\lambda 5050$ line because its Zeeman pattern is very similar that of Nd~{\sc iii} $\lambda 6145$.  In the stellar spectra this line has three clearly distinct groups of Zeeman components, and because the Land\'{e} splitting factors for the lower and upper levels are almost equal, the actual detailed splitting pattern does not deviate much from a simple triplet.  Thus Nd~{\sc iii} $\lambda 5050$ is an excellent line for measuring \bs\ in order to have a check on our measurements of the $\lambda 6145$ line. We were able to take advantage of this line because of the the large spectral windows covered by our new spectra.  

The atomic data required to convert splitting measurements into \bs\ values were taken from the Vienna Atomic Line Database (VALD) \citep{vald4,vald2,vald3}, except for some lines of Si, which were given Land\'{e} factors assuming L-S coupling.  To obtain \bs\ values, the equation
\begin{equation}
\bs = \frac{\Delta\lambda}{4.67 x 10^{-13}\lambda_{o}^{2}z},
\label{Bs-eqn}
\end{equation}
was used, where $\Delta \lambda$ is the shift of the $\sigma$ component from the zero field wavelength, $\lambda_{0}$ is the location of the spectral line with zero field, and $z$ is the effective Land\'{e} factor of the line (respectively 1.17, 1.00, and 1.35 for $\lambda 5050$, 6145, and 6149).  For a more complete discussion, refer to \citet{Mathys97} or \citet{summary1}. The goal was to reproduce the methods used by Mathys and collaborators as closely as possible so as to have \bs\ values that could be directly compared to theirs, in order to use the new measurements to improve the precision of the rotation period.  For the Fe~{\sc ii} line, the two sigma components of the Zeeman pattern are separated by as much as 1~\AA\ and are easily measurable.  The Nd~{\sc iii} lines have three components and the shifts between the central and blue-shifted components were measured to derive the mean field modulus.  The sigma components of both Nd~{\sc iii} lines are nearly symmetric, which allows for the measurement of \bs\ to be nearly approximation free.  This is consistent with the procedures outlined by \citet{MM2000}. The results of these measurements are summarized in Table \ref{spectra} in columns 6 through 8.  

To estimate uncertainties, multiple measurements for each spectrum were made by repeatedly fitting Gaussians to individual Zeeman components of measured lines, using the fitting capability of the IRAF {\it splot} function.  The location of the Zeeman components that determine $\Delta\lambda$ can be determined to within $\pm$0.01~\AA\ or less.   The corresponding $\Delta \lambda$ was then calculated for each measured line.  The measurements made from Equation (\ref{Bs-eqn}) are extremely sensitive, with deviations in $\Delta\lambda$ of the order of 0.01~\AA\ changing the measured \bs\ values by as much as 800~G (depending on the spectral line being measured).    

It is reassuring that the measurements of \bs\ for both Nd~{\sc iii} lines agree within the estimated uncertainties; again, this suggests that the uncertainties are realistic. We notice in the \bs\ measurements the same general phenomenon observed in the \bz\ measurements, namely that the field strengths as measured from a single spectrum using lines of different elements are not the same. The difference is not as striking for \bs\ as for \bz, but presumably points to the same phenomenon: the abundance distributions of the elements used, and perhaps the field structure,  are sufficiently inhomogeneous on a scale smaller than a stellar radius that different elements sample the field geometry in significantly different ways. However, it is also remarkable that the value of \bs\ varies through the rotation cycle by only about $\pm 10 - 15$\% from the  median value observed. 

\section{Refinement of the rotation period}

\citet{MM2000} used photometric data, and \bs\ measurements taken between August 1992 and September 1998, to test the rotation period of HD~318107 proposed by \citet{North1987}.  They found that the the 52.4~d period did not fit all their data. They showed that instead a period of $P = 9.7085 \pm 0.0021$~d fit both the photometry and \bs\ measurements.  When we use this period, we choose the zero point for phase to be JD 2449397.828 which is equivalent to the zero point used by \citet{MM2000}, but close to the mid-point of their \bs\ dataset.  We find that the phases of our recent data have a phase which is uncertain by approximately 0.1 cycle relative to the older measurements (this is the phase uncertainty computed from the one-sigma period uncertainty).  Our new \bz\ and \bs\ data, and the $I$ spectra we wish to model for abundances,  cannot be accurately phased with the older measurements without improving the precision of the period. 

In principle, the period can be improved both by fitting our new \bz\ measurements to the five measurements used by \citet{LM2000}, and by optimising the fit of our new \bs\ data to the well-determined variations of this field moment as measured by both the Fe~{\sc ii} and the Nd~{\sc iii} line. In practice, however, the variation of \bz\ with phase is too sparsely sampled by both the older field measurements and our own for comparison to allow us to refine the period. This problem is made more difficult for HD~318107 by the very different values of \bz\ that are found at any particular phase by only moderately different measurement methods.

Hence we turn to the new \bs\ data. Fortunately the numerous \bs\ values as measured by Fe~{\sc ii} 6149~\AA\ and by Nd~{\sc iii} 6145~\AA\ vary in significantly different ways, and from \citet{MM2000} we have convenient polynomials describing the variations of \bs\ with phase for these two elements.  Note that in Table~2 of \citet{MM2000} the coefficients $A$ of the polynomials for the two spectral lines have been inadvertently exchanged (private communication with Mathys). Varying the period through the range allowed by the uncertainty of $\pm 0.0021$~d, we find that our new \bs\ measurements fall very nicely on the appropriate polynomials for a limited range of periods. The best fit as judged by the $\chi^2$ of the fit for Fe~{\sc ii} is $P = 9.7087 \pm 0.0012$~d, while both Nd~{\sc iii} lines agree on $P = 9.7089 \pm 0.0007$~d, for a conservative global average of $P = 9.7088 \pm 0.0007$~d, and an adopted zero point of JD 2449397.828. This reduces the phase uncertainty of our new data with respect to the data from the 1990s to about $\pm 0.03$ cycle. All phases reported here are on this system. 

\begin{center}
\begin{figure*}
\centering
\includegraphics*[angle=-90, width=0.70\textwidth]{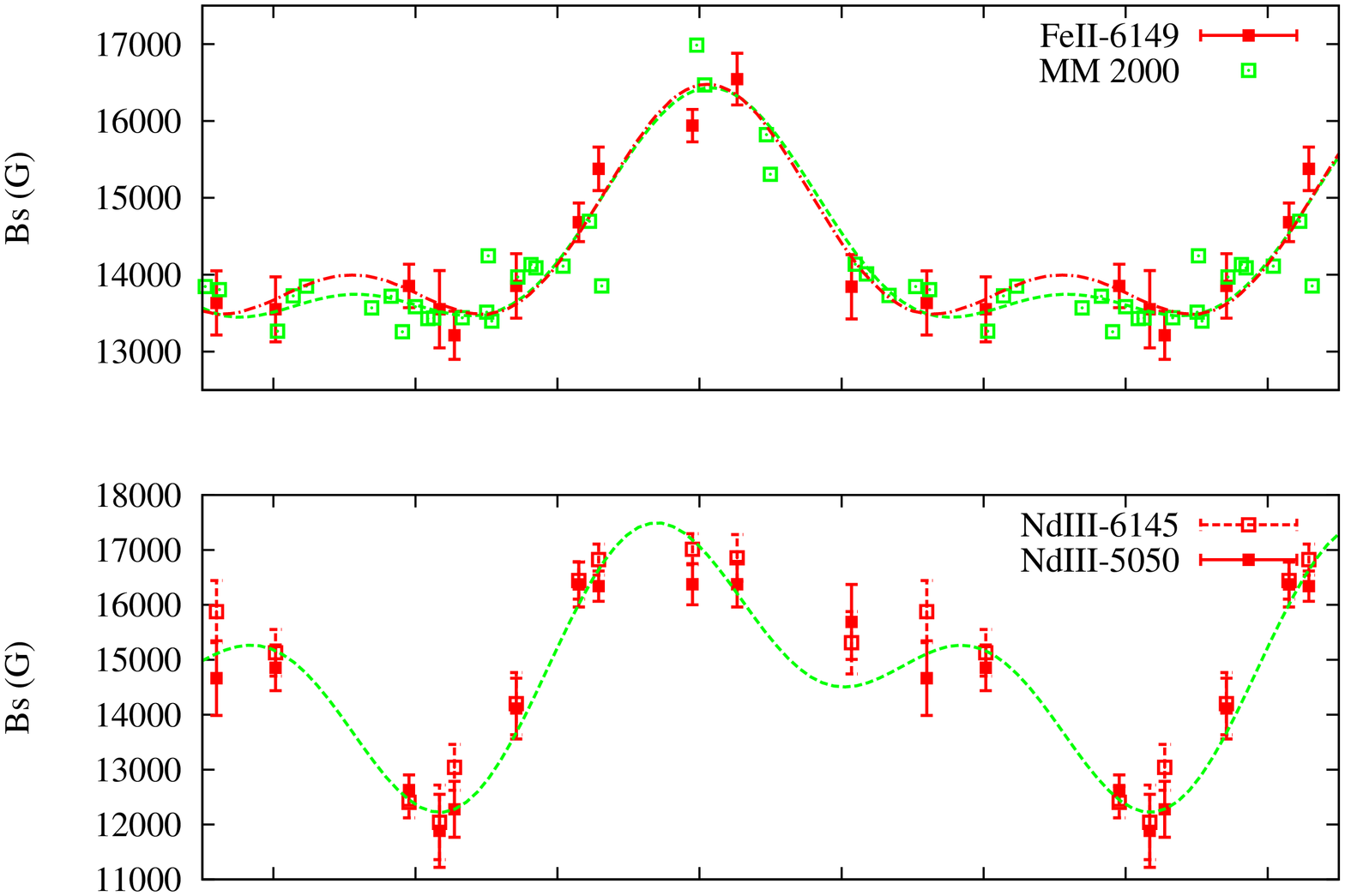}
\includegraphics*[angle=-90, width=0.70\textwidth]{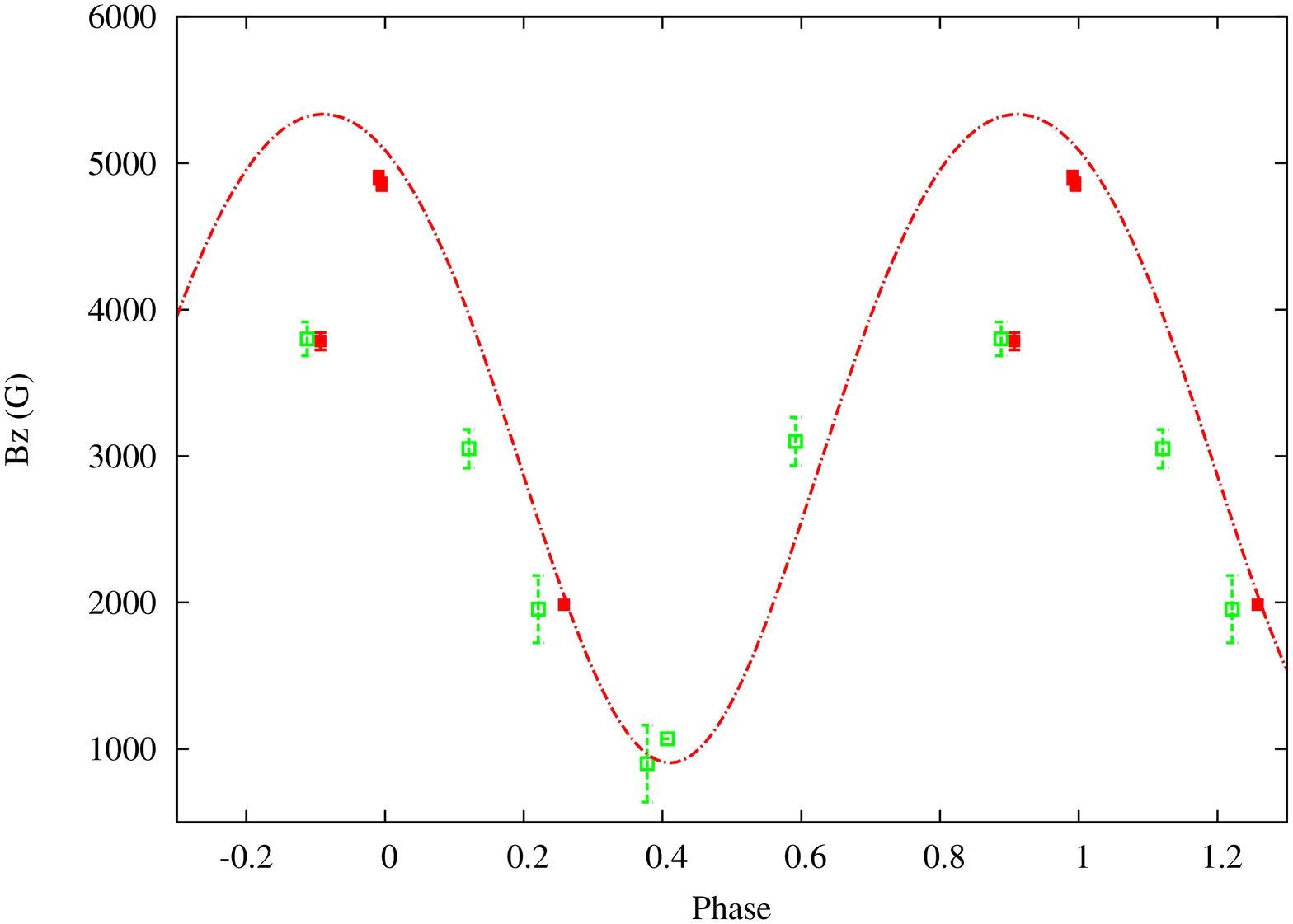}
\caption{The top two plots show our new \bs\ data phased with our best fit period, superimposed on the polynomial fits of \citet{MM2000} to their larger \bs\ dataset (green dashed curve).  In the top panel the second, slightly different, \bs\ curve (red dash-dotted line) is the variation of \bs\ predicted by our adopted field geometry.  The top panel data shows our individual measurements of \bs\ using Fe~{\sc ii} $\lambda 6149$ (data with error bars are ours, dots are from \citet{MM2000}); the second panel shows our measurements of \bs\ using Nd~{\sc iii} $\lambda 5050$  (filled symbols) and $\lambda 6145$ (open symbols).  The bottom figure depicts the \bz\ field variations observed in HD~318107.  The data points in green (open squares) are the measurements shown by \citet{LM2000}, shifted in phase to the present system; the red points (filled squares) are the new data from Table~\ref{spectra} plus (at phase 0.91) the metal line \bz\ value reported by \citet{paper1}. The dash-dotted line denotes the \bz\ rotational variation predicted by our adopted field geometry and phase system.}
\label{magfield}
\end{figure*}
\end{center}

A comparison of our new \bs\ data to the polynomials of \citet{MM2000} is shown in the upper two panels of Figure~\ref{magfield}.  It is clear from this figure that the agreement of our data with the previous \bs\ measurements is very satisfactory, and it is also obvious why significant phase shifts of especially the Nd~{\sc iii} data rapidly lead to decreased quality of the fit between the two data sets, so that the new period is accurately determined. 

The improved period is then used to phase the older \bz\ data shown by \citet{LM2000} with the new measurements. The results are shown in the bottom panel of Figure~\ref{magfield}. Several points emerge from this figure. First, the new data are generally consistent with the older measurements provided that we use the metal line \bz\ value from FORS1 at phase 0.91. At this phase the Balmer line field strength is about 2500~G higher than the older measurement at almost the same phase. Secondly, the variations of both \bs\ and \bz\ (from $\phi = 0.5$ to $\phi = 1.0$) appear to be rather non-sinusoidal. These facts, as discussed above, very strongly suggest that either the detailed metal abundance distributions or the field geometry, or probably both, are complex. This star  appears to belong to the small sample of stars, including HD~32633 \citep{hd32633}, HD~37776 \citep{Kochetal10}, HD~133880 \citep{Landstreet1990}, HD~137509 \citep{kochukhov2006}, and $\tau$~Sco \citep{donati2006} in which the magnetic field may depart in an important way from the generally dipolar topology of the magnetic fields of most Ap stars. (The field and surface He distribution of the outstandingly peculiar magnetic He-strong star HD~37776 have recently been mapped by \citet{Kochetal10}, and the resulting map reveals a quite astonishing degree of complexity.) Thirdly, the fact that the FORS1 metal line \bz\ value agrees with the older metal line measurement, while the Balmer line \bz\ measurement does not, is certainly consistent  with the idea that the difference between the two FORS1 \bz\ values is due to the combination of a rather complex field distribution and substantial differences in the way that the field is sampled by the presumably uniformly distributed H and the strongly patchy metal distributions. It is clear that it would be extremely interesting to obtain $V$ (and probably $Q$ and $U$ spectra) of this star with rather fine phase resolution, at least of order 0.05 cycle, to make it possible to explore more fully the field and abundance distribution structure.

\section{Modelling technique}

To model the spectrum of HD~318107 we use the magnetic spectrum synthesis programme ZEEMAN \citep{L1988,L1989}. This programme is designed to compute the emergent spectrum of a star of specified \te\ and $\log g$, with a specified magnetic field strength and geometry (currently characterized as a sum of colinear dipole, quadrupole, and octupole, at a specified angle $\beta$ to the rotation axis).  Either a uniform surface abundance distribution or a distribution which is a simple function only of latitude in the frame of reference of the magnetic axis can be specified. The parameters of the magnetic field model are chosen to match the observed field strength measurements of \bz\ and \bs\ as a function of phase, generally as described by \citet{LM2000}. The magnetic model can be further tested by comparison of computed line profiles to observed ones in cases such as HD~318107 where line splitting is an important line broadening mechanism. In principle, since ZEEMAN computes all four Stokes parameters, comparison could be made with observed polarised spectra, but this possibility is not implemented at present. 

The abundance variation with magnetic latitude is currently specified in the form of uniform abundance values on each of one to six rings of equal extent in latitude. The abundance of one element at a time, and its variation with latitude, can be varied by the programme to optimise the fit to a set of spectral line profiles in one or several spectra taken at various rotational phases. The output is then a coarse model of the abundance distribution of the element in magnetic latitude, effectively a very low resolution 1D map. This modelling is repeated for various elements. 

It is important to understand that the abundance model used by ZEEMAN was adopted to describe the kind of abundance variations observed in 53~Cam (HD~65339), a magnetic Ap star in which there are huge differences in large-scale mean abundances between the two hemispheres surrounding the two magnetic poles \citep{L1988}. In this case, a first order description that accounts only for abundance variation with magnetic co-latitude is an appropriate approximation. As will be discussed below, such large-scale variations does not seem to be a prominent feature of HD~318107, so this particular model is not a good approximation to the true abundance map. However, using it does appear to provide some useful information about the nature of the abundance distributions.

ZEEMAN includes a reasonably complete model of LTE line formation and radiative transfer in a magnetic field. The stellar atmosphere model is interpolated from a grid of precomputed solar abundance ATLAS 9 models. The four equations of radiative transfer for polarised light are solved on a grid with 0.01~\AA\ spacing. Local line profiles are computed as Voigt profiles based on atomic parameters taken from the VALD database or reasonable approximations. Line blending is correctly computed, by adding (polarised) line opacities due to various lines before solving the equations of transfer.  The correctness of many of the oscillator strengths used has been confirmed by synthesis of a large number of non-magnetic stars (for which the abundance of each element is essentially constant over the surface). Note that the modelling uses atmosphere models computed for solar abundance throughout; the atmosphere model is not updated as the abundance analysis proceeds. With rather large overabundances of some Fe peak elements found in this analysis, this fact certainly compromises the accuracy of the resulting abundance values \citep{KhaShu07}. 

The models of the field geometry (a low order axisymmetric multipole expansion) and of the distribution of elements over the surface (a few simple discrete rings axisymmetric about the magnetic axis) are quite schematic.  Compared to the sort of mapping done by \citet{Kochetal04,Kochetal10}, who map field vector and element distributions over the surface with $\sim 20^\circ$ spatial resolution in both surface coordinates, our models, in general, represent only a very rough approximation. However, the data set required for detailed mapping is far more extensive than the few, irregularly distributed, mostly $I$ spectra, obtained with a variety of resolving powers, and covering a large range in SNR, that are available at present for HD~318107. The type of modelling we carry out here is quite appropriate for the limited data available, and provides a first exploration of physical conditions and chemical abundances in the atmosphere of this star, which can be used for statistical studies in which each star is characterised by only a few numbers, and for identifying stars of such great interest that they warrant the extensive investment in observations needed for detailed mapping.

\subsection{Magnetic field model}

Before starting abundance analysis, we need to establish an approximate magnetic field model to use. An important quality of this model should be to reproduce reasonably well the phase variations of Zeeman splitting, so that the computed spectral line profiles are similar in form to the observed lines, and model fits are primarily sensitive to chemical abundance. 

To estimate the inclination $i$ of the rotation axis and the parameters of a suitable model magnetic field structure, we follow  \citet{Preston1967,Preston1970} and \citet{LM2000}. 

An estimate of the radius of the cluster member HD~318107 can be made from the values of $\log(\te)  = 4.072 \pm 0.02$ and $\log (L/L_{\odot}) = 1.92 \pm 0.1$ from \citet{paper2}.  The stellar radius is then found to be $R = (2.22 \pm 0.34) R_{\odot}$.  Our adopted rotation period of $P = 9.7088 \pm 0.0007$ days (Sect. 4) and $v \sin i = 7 \pm 2$ \kms\ (see below) may then be used in the equation  
\begin{equation}
\sin i =\frac{(v \sin i)P}{50.6R},
\label{raid}
\end{equation} 
where $v \sin i$ is in \kms\ and $R$ is in solar units, to obtain $i = 37^\circ \pm 15^\circ$. This is not a very strong constraint, but provides a useful check to the value of $i = 22^\circ$ derived below.  

From the form of the \bz\ and \bs\ variations with rotational phase, we can get a rough idea of the parameters the multipole model of the field will require. The fact that both magnetic moments (especially \bz) vary significantly with phase indicates that $i$ must be substantially different from zero, as we found above. The fact that all the \bz\  data have the same sign indicates that, as the line of sight executes a cone around the rotation axis on the star, it always remains within one magnetic hemisphere, so $i + \beta < 90^\circ$. We expect that the sum of the polar fields will be of order 15~kG. From the phase behaviour of \bz, it appears that the closest approach of the line of sight to the magnetic axis occurs at about phase 0.9, even though the value of \bs\ is near minimum there; apparently the local field strength increases fairly rapidly into the magnetic hemisphere that is observed only briefly around phase 0.4. 

To obtain a detailed model the FORTRAN programme FLDSRCH \citep{LM2000} was used.  The programme takes as input values of \bz\ and \bs\ at four equidistant phases. The relative importance of these two kinds of data can be varied by adjustable weights. Since we are particularly concerned with having a field model that predicts the correct amount of Zeeman splitting in lines at various phases, we have given a relatively large weight to the \bs\ data. 
FLDSRCH calculates \bz\ and \bs\ as functions of phase for a field composed of colinear dipole ($B_{\rm d}$), quadrupole ($B_{\rm q}$), and octopole ($B_{\rm oct}$) components.  The programme iteratively searches for the values of $i$, $\beta$, $B_{\rm d}$, $B_{\rm q}$, and $B_{\rm oct}$ that best fit the values  \bz\ and \bs\  provided as input.  Along with the best fit parameters, the programme provides calculated values of \bz\ and \bs\ as functions of phase. 

The best fit model found in this way (with of course some substantial arbitrariness because the data are clearly more complex than can be reproduced by the simple magnetic field model adopted) are 
 $i = 22^{\circ}$, $\beta = 65^{\circ}$, and polar field strengths of $B_{\rm d} = +25600$~G, $B_{\rm q} = -12800$~G, and $B_{\rm oct} = +900$~G.  The $\bz(\phi)$ and $\bs(\phi)$ predicted by this model are plotted in Figure \ref{magfield}, and are in reasonably good agreement with observations of \bs\ but reproduce only the general amplitude of the \bz\ variations.  This model is a poor representation to the actual field structure, thus specifying uncertainties would give an unrealistic impression of how accurate this model is.  However, experimentation with fits to the variations of \bz\ and \bs\ suggest that we can best reproduce the observed data when the model parameters are in this region of parameter space.

\begin{center}
\begin{figure*}
\centering
\includegraphics[width=0.45\textwidth]{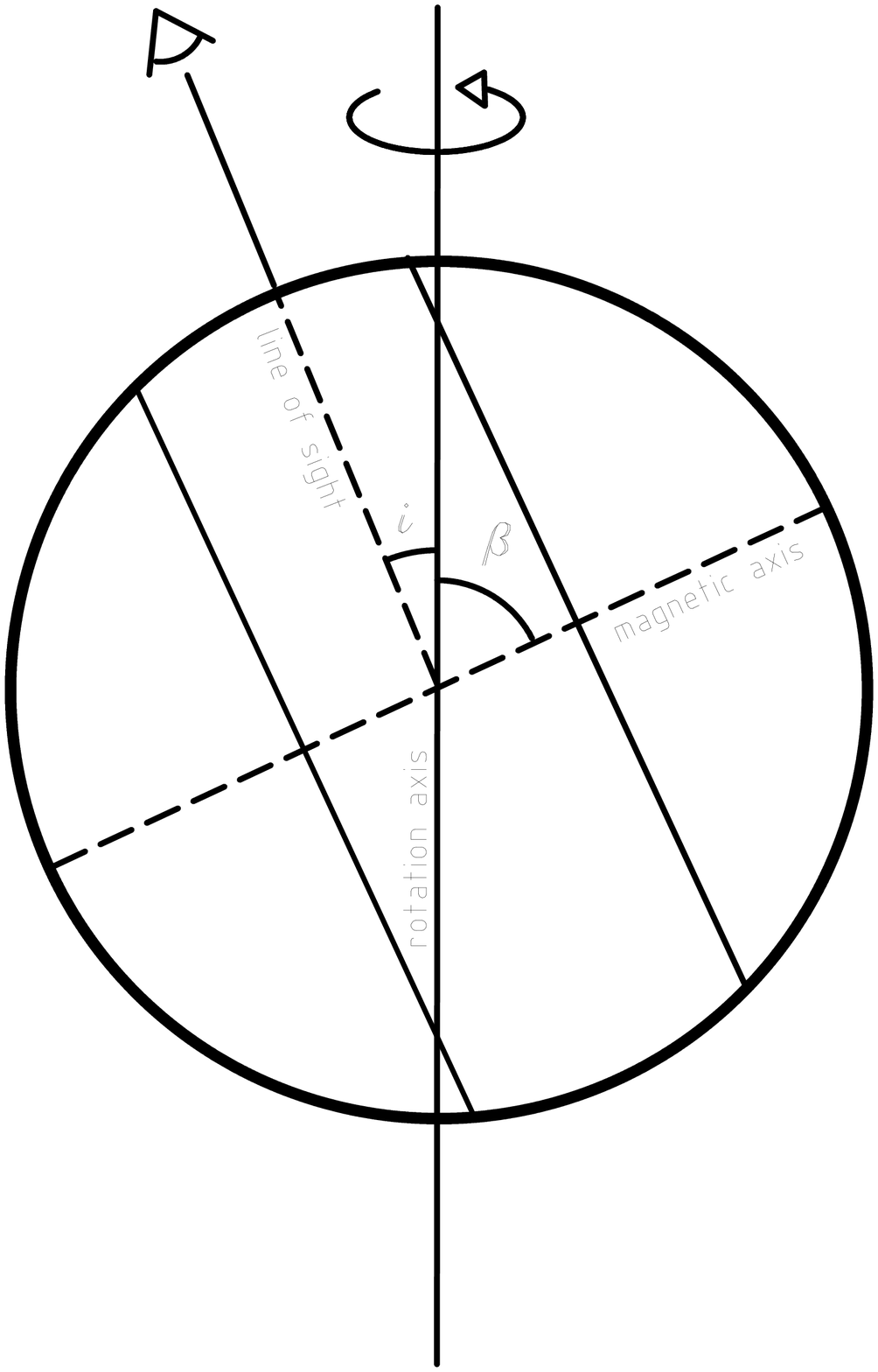}
\includegraphics[width=0.45\textwidth]{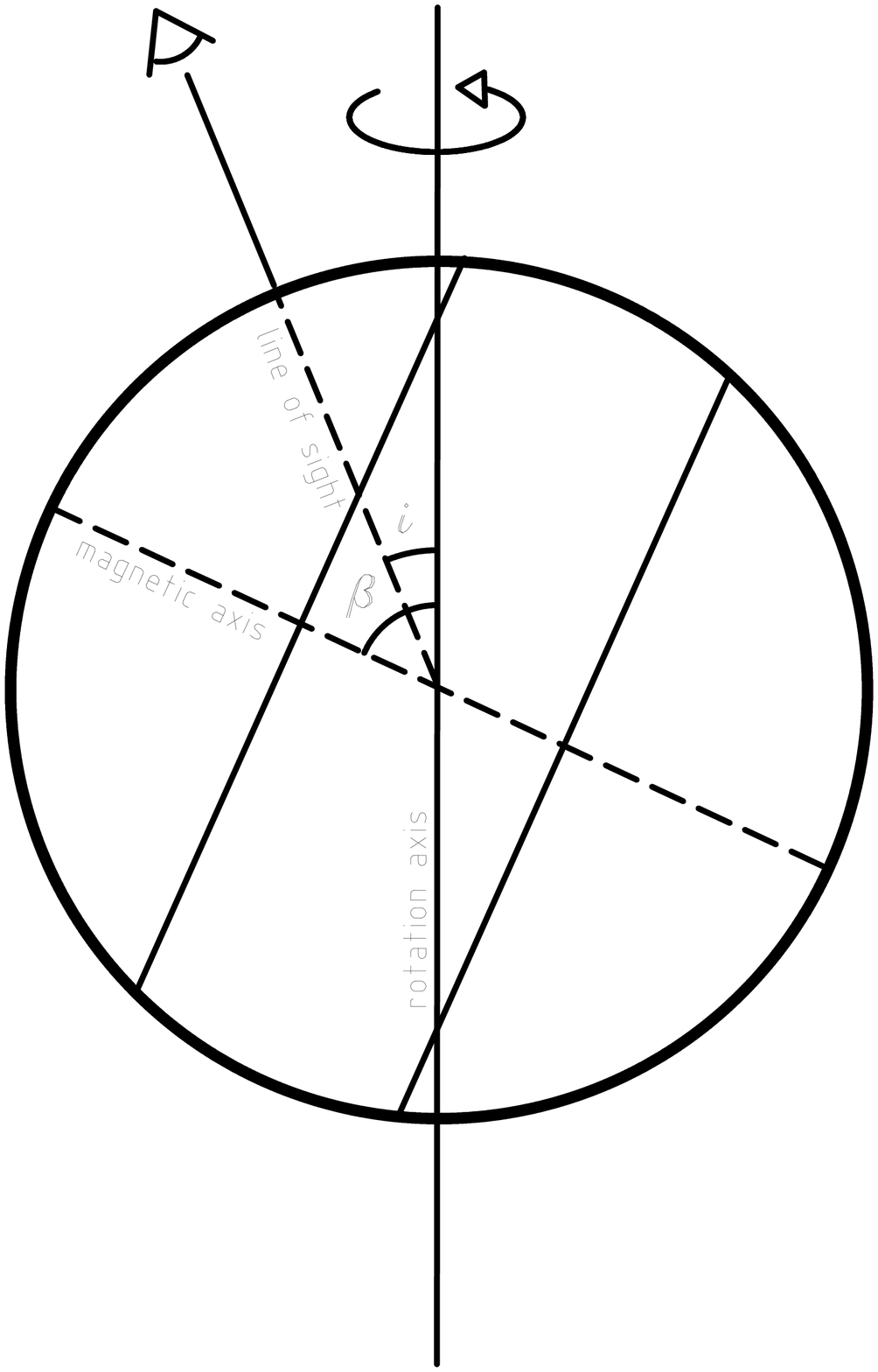}
\caption{The geometry adopted for HD~318107.  The vertical axis is the rotation axis of the star.  The angle between the line of sight and the rotation axis is $i = 22^{\circ}$.  The angle between the rotation axis and the magnetic field axis is $\beta = 65^{\circ}$.  For HD~318107, $i + \beta < 90^{\circ}$ and therefore, as the star rotates, the negative magnetic pole is never directly observed.  The left panel depicts an orientation where the magnetic axis is farthest from the line of sight ($\phi = 0.4$) and the right panel is when the magnetic axis is nearer alignment with the line of sight ($\phi = 0.9$).  The bands used for dividing the abundance distribution are shown as solid lines and are normal to the magnetic field axis.}
\label{oblique}
\end{figure*}
\end{center}

The resulting geometry is shown in Figure \ref{oblique}.  The figure is a graphical representation of the oblique rotator model that displays the relationship of the line of sight to the two angles $i$ and $\beta$ and to the field axis for the phases 0.4 (left figure) and 0.9 (right), when the line of sight is respectively farthest from and nearest to the visible magnetic pole. 

\subsection{Chemical abundance model}

With the stellar parameters and a model of the magnetic field established, the elemental abundance distributions are obtained using the synthesis program ZEEMAN described above.   ZEEMAN compares the calculated line profiles to the observed profiles and varies the abundance to minimize the reduced $\chi^{2}$ of the difference between the model and the observed $I$ spectral lines.  The programme automatically adjusts \vsi\ and the radial velocity $v_{\rm R}$ to optimal values at each iteration. For the analysis here, \te\ was set to $11\,800$~K, and  $\log g$ to 4.2, consistent with the value $4.22 \pm 0.13$ determined from $M = 2.95 \pm 0.15 M_\odot$ \citep{paper2} and the radius found above. Note that because only one ionisation stage is detected for almost all elements, and because the shapes of spectral lines are fit only approximately at best, we do not obtain very strong constraints on the value of \te\ from modelling the observed spectrum.  However, we note that our adopted \te\ is consistent with the model fits of \citet{korucz1979} when comparing the computed and observed H$\beta$ profiles.  

Atomic data were taken from the VALD.  The best-fitting value of $v \sin i$ was found to be $7 \pm 2$~\kms.  The best fit required some experimentation and visual evaluation, as the \vsi\ broadening is smaller than the Zeeman splitting, so the optimal value is best found by looking at how various \vsi\ values blend the Zeeman components together. The microturbulence parameter was set to 0~\kms, as the strong magnetic field is expected to completely suppress the convective motions underlying this line broadening agent. 

With the twelve spectra listed in Table~\ref{spectra}, we have fairly good phase coverage of the rotation cycle of HD~318107. The largest gap is between phases $\phi = 0.80$ and 0.99. However, some of the HARPS spectra have quite low SNR. Considering that none of the spectral lines seem to vary very much in equivalent width, it appears that an adequate sampling of the rotation cycle for this first exploratory modelling would be to start with a considerably smaller number of spectra, focussing on the best available ones. We have chosen to model only five spectra: the two ESPaDOnS spectra at $\phi = 0.995$ and 0.258, the UVES spectrum at $\phi = 0.390$, and two HARPS spectra at $\phi = 0.615$ and 0.803. For lines below 4795~\AA, we replace the UVES spectrum with the lower resolution FLAMES-UVES spectrum ($\phi = 0.453$). With this choice of spectra to model, we are using the spectra with the best resolving power and SNR, we have roughly uniform spacing of about $\Delta \phi \sim 0.2$ in phase (and therefore sample coarsely the full range of available directions of the line of sight relative to the magnetic axis), and we reduce the computation time considerably. 

Spectral lines respond non-linearly to changes in abundance. Thus fitting several lines of different strengths provides a stronger test of the correctness of the model than fitting one single line. 
Where possible,  the best-fit abundance distributions were found by simultaneously fitting multiple unblended lines of a given element within a window of a few 10s of Angstroms.  The spectral lines modelled are discussed in the sections on individual elements. 

HD~318107 does not fulfill the conditions for which Zeeman was designed (see Sect. 4).  The line of sight remains firmly in a single hemisphere, and in fact, because the $i$ value of our model is small, the line of sight only varies in magnetic co-latitude over a total range of about 45$^\circ$. Furthermore, the strength of most spectral lines vary with rotational phase only rather modestly, suggesting that the pole-to-pole variations of abundance may not be very large. 

Nevertheless, we experimented extensively with multi-ring models (i.e. we allowed the abundance to vary with magnetic latitude when we searched for the best fit abundance model for each element). We found that a maximum of three rings should be used (abundances constant on rings 60$^\circ$ wide measured from the visible magnetic pole) in order to achieve convergence to the same model regardless of the initial abundances assumed. When we ran 3-ring models using different spectral regions (and thus different sets of spectral lines), we found for a few elements (particularly Si and Ti) that the best fit was for a model with substantial differences ($\sim 1$~dex) between one ring and another, but no consistent pattern emerged from modelling different spectral regions. We eventually concluded that the best fit models with abundances which seem to vary strongly with magnetic co-latitude were probably actually fitting some other kind of abundance inhomogeneity, probably on a smaller scale, as already strongly suggested by the very different values of \bz\ found at a given phase using lines of different elements (cf. Table~2). 

Our conclusion was thus that the models found by ZEEMAN with strong variation in abundance with magnetic co-latitude are unphysical. In fact there is no convincing evidence for strong abundance variations on a global scale. We have therefore adopted a much simpler approach to abundance modelling. The abundance of each element studied is assumed to be uniform over the stellar surface. The best fits we obtain to groups of lines in various spectral windows of our chosen five spectra, sampling the rotation cycle of the star, of course do not yield the same mean abundance in all windows. We use the standard deviation of the abundances found in different windows as a simple measure of the uncertainty of the value of the best fit mean abundance. It will be seen that the uniform abundance values determined for different spectral windows are roughly concordant, and that with this simple assumption we can fit all spectral lines modelled fairly well at all phases, giving us some confidence that the derived average abundances are meaningful.

\section{Chemical abundances of individual elements}

\begin{center}
\begin{figure*}
\centering
\includegraphics[angle=0,width=0.95\textwidth]{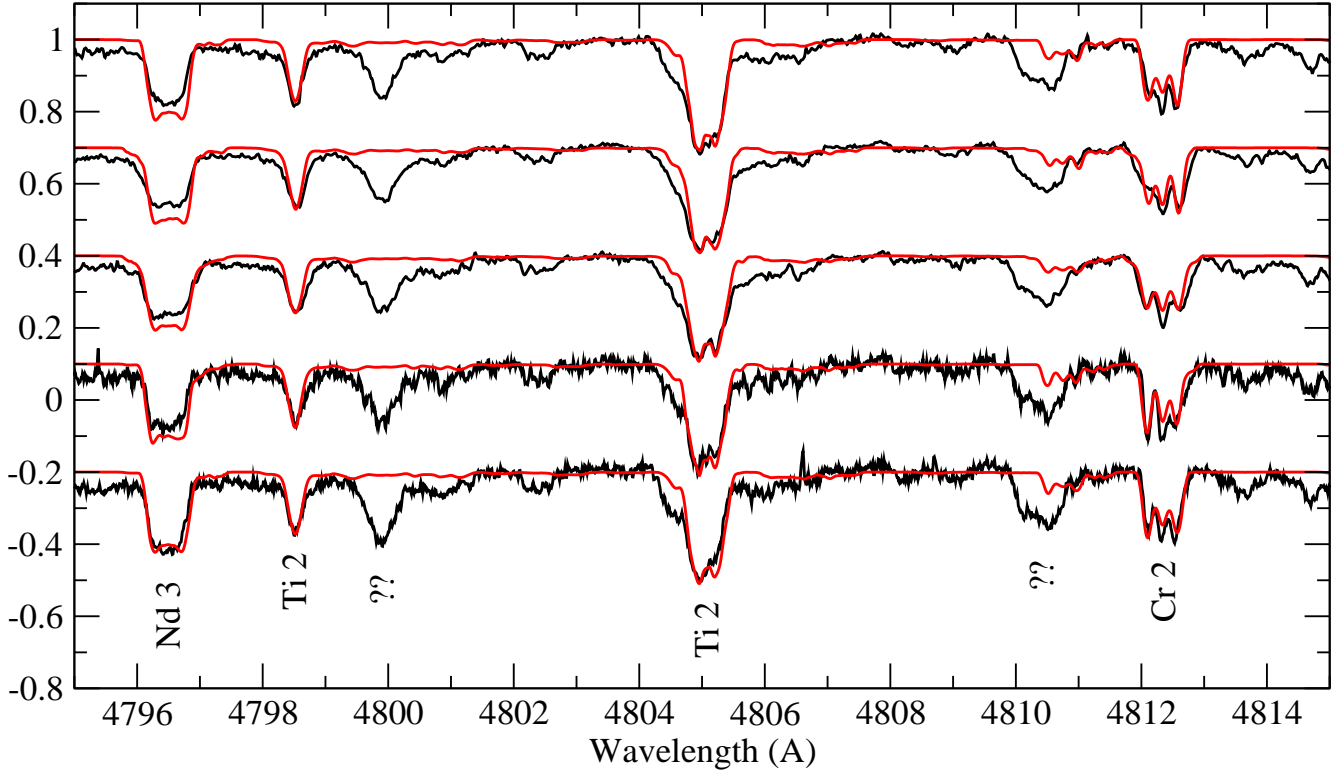}
\caption{Spectrum synthesis of the region 4795 -- 4815~\AA using mean abundances. Black curves are observed spectra, red curves are models. From top to bottom phases 0.991, 0.258 (both ESPaDOnS), 0.390 (UVES), and  0.615 and 0.803 (both HARPS). The line of sight is closest to the magnetic pole at phase 0.9, and to the magnetic equator at phase 0.4. Note that some strong lines are still unidentified. }
\label{spec48}
\end{figure*}
\end{center}   

\begin{center}
\begin{figure*}
\centering
\includegraphics[angle=0,width=0.95\textwidth]{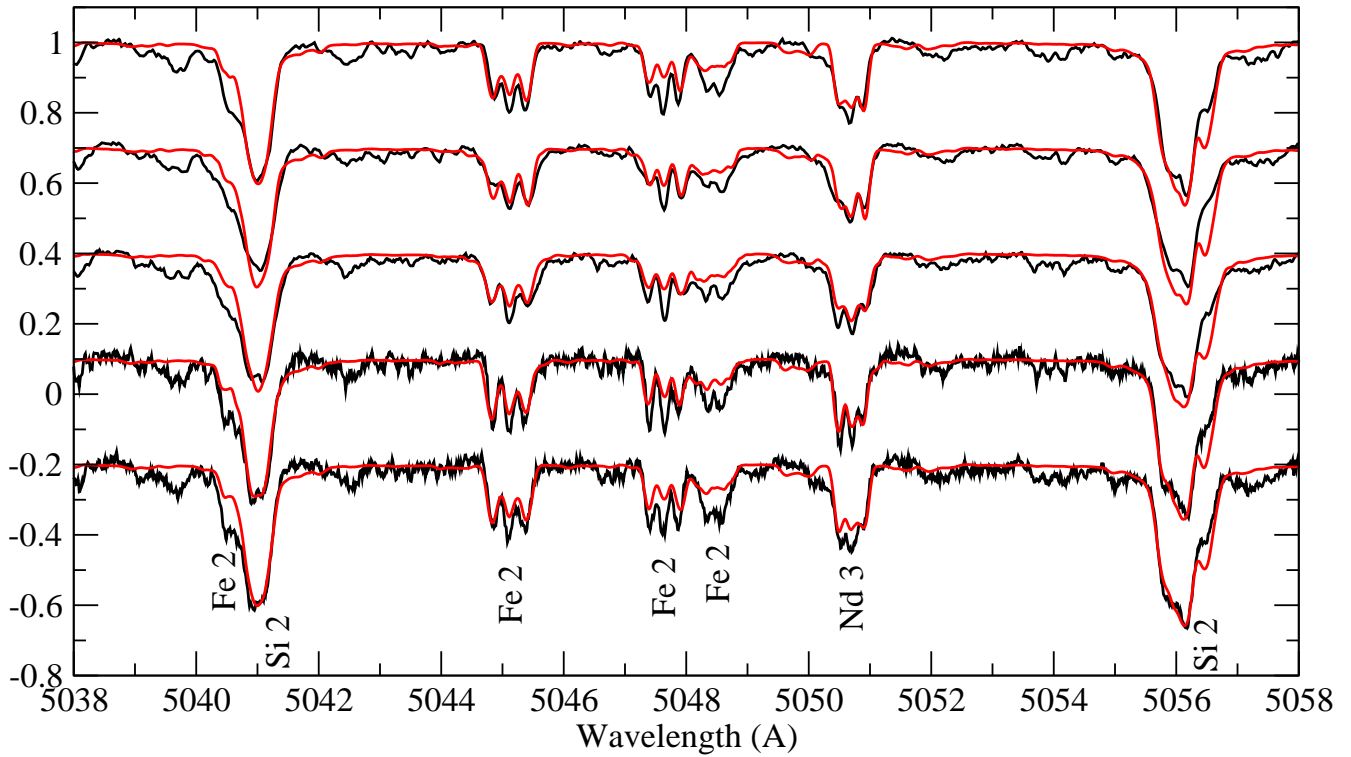}
\caption{Spectrum synthesis of the region 5038 -- 5058~\AA using mean abundances. Black curves are observed spectra, red curves are models. From top to bottom phases 0.991, 0.258 (both ESPaDOnS), 0.390 (UVES), and  0.615 and 0.803 (both HARPS). The line of sight is closest to the magnetic pole at phase 0.9, and to the magnetic equator at phase 0.4.}
\label{spec50}
\end{figure*}
\end{center}   

Because of the large wavelength coverage of the available spectra of HD~318107, our data are sufficient to obtain first approximations to the chemical abundance distribution for O, Mg, Si, Ca, Ti, Cr, Fe, Nd, and Pr, and useful upper limits for the abundance of He and Mn.  The large wavelength coverage also means that multiple lines of many elements can be found that have a variety of strengths and splitting patterns.

The mean abundance values found for the elements studied are listed in Table~\ref{avgabund}, together with the uncertainties estimated from the scatter of fits to (usually two) different spectral windows. For elements for which only a single line or region was available, we have estimated the uncertainty as $\pm 0.2$~dex. For comparison, the solar abundances reported by \citet{Aspletal09} are tabulated as well. The quality of the fits is illustrated with short spectral windows, showing the comparison of uniform abundance models with the five observed $I$ spectra used for the fitting, in Figures~\ref{spec48} and \ref{spec50}. In each figure the model spectra have been computed with the global mean from Table~\ref{avgabund} rather than the local best fits. The lines used and the consistency of the fits are discussed element by element below. 

\begin{table}
\caption{Mean abundances of elements studied}
\begin{tabular}{lrr}
\hline\hline 
Element   &   $\log(n_{\rm X}/n_{\rm H}) \pm \sigma$  &  $\log(n_{\rm X}/n_{\rm H})_\odot$ \\
\hline 
He           &   $\leq -2.50$          &     $-1.07$ \\
O             &   $-3.65 \pm 0.2$      &  $-3.31$ \\
Mg           &   $-5.70 \pm 0.2$      &  $-4.40$ \\
Si            &   $-3.65 \pm 0.15$     &  $-4.49$ \\
Ca           &   $-5.55 \pm 0.2$      &  $-5.66$ \\
Ti            &   $-5.05 \pm 0.1$      &   $-7.05$ \\
Cr           &   $-4.50 \pm 0.15$     &   $-6.36$ \\
Mn          &   $\leq -6.00$             &   $-6.57$ \\
Fe           &   $-3.00 \pm 0.2$      &   $-4.50$ \\
Pr            &   $-6.35 \pm 0.2$      &   $-11.28$ \\
Nd           &   $-6.20 \pm 0.15$    &   $-10.58$ \\
\hline\hline \\
\end{tabular}
\label{avgabund}
\end{table}

\subsection{Helium}

The strongest lines of He~{\sc i} in the available spectra should be the triplet lines at 4471 and 5876~\AA. Neither of these lines is unambiguously detected, and fitting these regions provides an upper limit to the abundance of helium of at least a factor of 20 below the solar abundance. HD~318107 is definitely a He-weak star. 

\subsection{Oxygen}

The nine weak O~{\sc i} lines at 6155-56-58~\AA\ are blended with lines of Pr, and are too weak to model usefully. In contrast, the intrinsically stronger lines at 7771-74-75~\AA\ are clearly visible and appear to be unblended. However, this spectral region is only available in the ESPaDOnS spectra, and so we have determined the mean abundance using only two spectra. The blend of Zeeman-split features is rather well fit in the two spectra about 0.25 cycle apart, and apart from possible non-LTE effects in these lines, the O abundance is unambiguously determined, and appears to be about 0.3~dex lower than the solar value. 

\subsection{Magnesium}

The only clean line found of sufficient strength to model is Mg~{\sc ii} $\lambda 4481$. This line is reasonably well fit at all phases with a uniform abundance, and indicates that Mg is about a factor of 20 lower in abundance than in the Sun. 

\subsection{Silicon}

In the available spectra, the cleanest reasonably strong lines are the single line at 5041~\AA, the doublet at 5055-56~\AA, and the single lines at 5955 and 5978~\AA. The fit to the two longer wavelength lines is quite good at all phases with a single mean abundance. The fit to the shorter wavelength lines is a little less satisfactory, as will be seen in Figure~\ref{spec50}. The fit to these lines is roughly equally good at all phases, suggesting that no large-scale abundance variation from hemisphere to hemisphere is detected. 

\subsection{Calcium}

Useful lines of Ca~{\sc ii} are found in the ESPaDOnS and HARPS spectra, in the near ultraviolet at 3933~\AA\ and in the ESPaDOnS infrared spectra at 8498, 8542, and 8662~\AA. The infrared lines are all blended with Paschen lines, and so the abundance of Ca was determined using only the $\lambda$3933 line in the two ESPaDOnS spectra. The abundance of Ca appears to be approximately solar. A weak line of Ca~{\sc ii} nearly coincides with a strong unidentified feature at 4799.9~\AA, but the computed strength of this line using the abundance derived from $\lambda 3933$ is negligible (see Figure~\ref{spec48}). 

\subsection{Titanium}

Lines of Ti occur throughout the blue spectral region available in almost all the spectra. We have modelled lines at 4468, 4488, 4518~\AA, and also lines at 4798 and 4805~\AA. The two lines in the longer wavelength window are both well fit at all phases (see Figure~\ref{spec48}), but in the shorter wavelength window the computed line is somewhat too strong (and not quite the correct shape) for the strong line at 4468~\AA\ and somewhat too weak for the weak line at 4518~\AA, which might be blended with another line not included in the line list. Nevertheless, synthesis of the two spectral regions yield very similar mean abundances, showing that Ti is overabundant with respect to the solar abundance by approximately 2~dex. 

\subsection{Chromium}

The abundance of Cr has been derived from modelling lines of multiplets 39 and 44 at 4539, 4555, 4558 and 4565~\AA, and lines of multiplet 30 at 4812 and 4824~\AA. All lines are fit reasonably well with a uniform abundance. The various Zeeman split line profiles are approximately reproduced (see Figure~\ref{spec48}). An overabundance of almost 2~dex with respect to the solar abundance is found. This result is roughly consistent with the variations of abundance with \te\ found in Ap stars by \citet{Ryabetal04}.

\subsection{Manganese}

The strongest visible lines of Mn are in the region below the Balmer jump, between 3430 and 3500~\AA. Since these lines are not included in any of the available spectra, we modelled several lines of Mn~{\sc ii} between 6122 and 6130~\AA. These high-excitation lines are not detected, so they only provide an upper limit of about 0.6~dex above solar abundance. It is interesting that although the even-Z iron peak elements Ti, Cr and Fe are overabundant by 1--2~dex, this does not seem to be the case for Mn. 

\subsection{Iron}

Iron has many usefully strong lines scattered through the spectrum. We have fit lines in three windows, 4488 -- 4542, 4820 -- 4830, and 5026 -- 5037~\AA. All the numerous Fe~{\sc ii} lines in these regions are reasonably well fit with an abundance about 1.5~dex above the solar abundance. The weaker lines seen in Figure~\ref{spec50} suggest that the abundance of Fe might be even a little larger than this value; in any case, they show the tendency, also found for Ti, that computed weak lines to be weaker than the observed lines. This result is roughly with the variation of Fe abundance with \te\ found in Ap stars by \citet{Ryabetal04}.

\subsection{Praseodymium}

Four useful lines of Pr~{\sc iii} are found in the region 6098 -- 6196~\AA.   All are fit reasonably well with a mean abundance almost 5~dex larger than the solar value.

\subsection{Neodymium}

Nd~{\sc iii} has lines scattered through the visible window. We have modelled lines at 4796, 4822 (Figure~\ref{spec48}), and 5050~\AA (Figure~\ref{spec50}), and find a mean abundance about 4.3~dex larger than the solar abundance. 

\section{Discussion \& conclusions}

This paper is the first in what is expected to be a series of reports describing modelling of stars from the survey of magnetic Ap stars in open clusters of known age. The aim of this project is to establish a preliminary model of the magnetic field structure, to estimate the chemical abundances of a number of common elements, and to extract whatever information is available in our data about the surface homogeneity or inhomogeneity of those elements by means of simple, rather schematic models whose parameters are derived by fitting synthesised spectra to observations. 

HD~318107 is a relatively hot magnetic Ap star, with $\te = 11\,800$~K. It has $\log L/L_\odot = 1.92$, and mass $M/M_\odot = 2.95$. The rotation period of the star is 9.7088~d, and it has $\vsi = 7$~\kms. From its membership in NGC~6405 we know that its age is $\log t = 7.80$ and that it has completed about 17\% of its main sequence lifetime. The star has a strong magnetic field, of order 15~kG globally.

We have used previously published (and modelled) magnetic field measurements of \citet{Mathys97},  \citet{MM2000},  \citet{LM2000}, and \citet{paper1}, together with measurements of our own, to improve the rotation period of the star to $P = 9.7088 \pm 0.0007$~d, which makes it possible to phase recent magnetic data and spectroscopy accurately with previous measurements, and thus to use the full set of magnetic and spectral data available to us to model the star. 

Our model of the magnetic field is a simple, low-order axisymmetric multipole expansion which retains the global topology of a dipolar field structure. This model is used in the framework of the oblique rigid rotator model to describe the observed variations of \bz\ and \bs\ with rotational phase. Fitting the predicted \bs\ and \bz\ variations approximately to the observed variations allows us to determine the parameters of the model magnetic field as well as the inclination $i$ of the rotation axis and the obliquity $\beta$ of the magnetic field to the rotation axis. Because the variations of the field moments are rather complex, and in fact poorly determined for \bz, the adopted model is only a rather coarse first approximation to the actual magnetic field geometry of HD~318107. However, this model is appropriately simple considering the limited information available about the star, in particular the small number of \bz\ measurements and high-dispersion polarimetric spectra available. 

The qualitative form of the field structure is clear from the nature of the field data shown in Figure~\ref{magfield}. The fact that \bz\ does not reverse sign indicates that the line of sight to the star is confined to one magnetic hemisphere, which in turn means that the sum $i + \beta \lesssim 90^\circ$. The fact that the minimum value of $|\bz|$ is considerably smaller than the maximum value shows that the line of sight does approach the magnetic equator, while the large ratio of the maximum value of $|\bz|$ relative to the value of \bs\ indicates that the line of sight does approach the magnetic pole. The actual parameter values adopted are discussed in Sec 4.1. The resulting magnetic model does not reproduce particularly well the detailed shape of the variations of \bz, but it works surprisingly well as a description of the Zeeman broadening (the dominant source of broadening) of the spectral lines we analysed. 

With only a dozen $I$ spectra scattered through the rotation cycle (but including spectra obtained near times of maximum and minimum \bz; i.e. when the line of sight is near the magnetic pole and near the magnetic equator), it is not possible to do detailed mapping. However, a preliminary reconnaissance of the abundances of several elements, and a first estimation of their variation over the surface, is possible. We have considered a model that uses three co-axial rings, one around each magnetic pole and one around the equator, but there is no strong evidence for strong abundance variations on the global scale.  The model finally adopted assumes uniform abundance over the stellar surface. This model provides the simplest possible quantitative description of abundance variations on the star. 

Assuming uniform abundances over the stellar surface, we find, as expected, that most elements have distinctly non-solar abundances. The abundance of O is slightly below solar, while Mg appears to be significantly lower than the solar abundances. Ca has nearly solar abundance.  Upper limits to the abundances of He and Mn are possible, the former unambiguously classifying HD~318107 has a He-weak star and the latter suggesting an abundance no more than 0.6 dex above the solar value.  The other studied elements (Si, Ti, Cr, Fe, Pr, and Nd) are all present with higher abundance than in the Sun.   For all of the elements studied, the abundances measured over the part of the surface visible in our spectra do not appear to vary strongly with magnetic co-latitude. However, for some elements a fit to the five spectra modelled is significantly better using the three-ring model than with a uniform abundance; we interpret this result as reflecting substantial small-scale abundance variations, as already hinted by the very different \bz\ values measured in the ESPaDOnS spectra using lines of different elements. 

HD~318107 appears to be a star which may merit further study. With its extremely large and apparently rather complex field, and large (and apparently fairly non-uniform) abundance anomalies, this may be a good candidate for more detailed mapping, which would require a major observing campaign to obtain phase coverage in all the Stokes components. The main difficulties of such a study are the lack of much Doppler broadening of the lines to provide spatial resolution in azimuth, and the small value of $i$ which limits the line of sight to one magnetic hemisphere. But even before this is done, obtaining a few more spectropolarimetric $V$ observations should help to clarify the nature of the variation of \bz\ with phase, enable more detailed modelling of the field geometry, and help us to understand the discrepancies between the values of \bz\ measured with Balmer lines and with various metals. 

\begin{acknowledgements}
JDB, JDL, JS, and GAW are grateful for support by the Natural Sciences and Engineering Research Council of Canada.  GAW acknowledges support from the DND (Canada) Academic Research Program.  Astronomy research at the Open University is supported by an STFC rolling grant (LF).  OK is a Royal Swedish Academy of Sciences Research Fellow supported by grants from the Knut and Alice Wallenberg Foundation and the Swedish Research Council.  CP acknowledges the support of the project P19503-N16 of the Austrian Science Fund (FWF).  
\end{acknowledgements}

\bibliographystyle{aa}
\bibliography{apstars}
\end{document}